\documentclass[preprintnumbers,amsmath,11pt,amssymb,floatfix,superscriptaddress,nofootinbib]{article}
\usepackage{array}
\usepackage{relsize}
\usepackage{cite}
\usepackage{amsmath}
\usepackage{amssymb}
\usepackage{graphicx}
\def\mytitle#1{\setcounter{equation}{0}
\setcounter{footnote}{0}
\begin{flushleft}\Large\textbf{#1}\end{flushleft}
\vspace{0.25cm}}
\def\myname#1{\leftline{{\large #1}}\vspace{-0.13cm}}
\def\myplace#1#2{\small\begin{flushleft}\textit{#1}\\
\texttt{#2}\end{flushleft}}

\usepackage{graphicx}
\begin{document}

\mytitle{Fate of an Accretion Disc around a Black Hole when both the Viscosity and Dark Energy is Effecting}

\myname{Sandip Dutta \footnote{duttasandip.mathematics@gmail.com} and Ritabrata Biswas\footnote{biswas.ritabrata@gmail.com}}
\myplace{Department of Mathematics, The University of Burdwan, Burdwan, West bengal, India-713104.}{} 

\begin{abstract}
This paper deals with the viscous accretion flow of modified Chaplygin gas towards a black hole as the central gravitating object. Modified Chaplygin gas is a particular type of dark energy model which mimics of radiation era to phantom era depending on the different values of its parameters. We compare the drak energy accretion with the flow of adiabatic gas. An accretion disc flow around a black hole is an example of a transonic flow. To make the model, we consider three components of Navier Stokes' equation, the equation of continuity and the modified Chaplygin gas's equation of state. As a transonic flow passes through the sonic point, the velocity gradient being apparently singular there, gives rise to two flow branches : one infalling, the accretion and the other outgoing, the wind. We show that the wind curve is stronger and wind speed reaches to that of light at a finite distance from the black hole when dark energy is considered. Besides, if we increase the viscosity, accretion disc is being shortened. This two process acting togather deviates much from adiabatic accretion case. It shows a weakening process for the accretion procedure by the works of viscous system influenced angular momentum transport and the repulsive force of the modified Chaplygin gas together.\\

\end{abstract}

\section{Introduction}
~~~~~Accretion, by word, means enriching up. Astrophysical accretion, either in a protoplanetary disc or around a  massive central engine, means fall of mass towards a gravitating center. Present article discusses the flow of diffused material towards a particular type of compact object, namely Black Hole(BH). Though firstly was a theoretical prediction, BHs, are now-a-days almost proved to be present in X-ray binaries, center of galaxies, especially in Active Galactic Nucleis (AGNs) or in several sources of gravitational waves \cite{Gravitational wave}. However, most the evidences are indirect, i.e., the critical behaviour of infalling matter towards a steep gravitational well brings us the information. Due to the conservation of angular momentum, the infalling flow towards a BH forms a disc like structure. In the disc, accreting fluid simultaneously obeys
$\frac{\partial}{\partial r}\left(\Omega R^2 \right)>0$ and $\frac{\partial \Omega}{\partial R}<0$,
where $\Omega$ is local angular velocity for Keplerian motion (i.e., 
$\Omega ^2 R=\frac{GM}{R^2}$ which implies $\Omega = \left( \frac{GM}{R^3}\right)^\frac{1}{2}$), R is the radial distance from the center.
Physically which means that angular momentum increases and angular velocity decreases outwards. This gives birth of the outward angular momentum transport. The outward force due to angular momentum is of order $\frac{1}{R^3}$ and inward gravitational pull is of order $\frac{1}{R^2}$, so when going towards a compact object, there is a region where gravitational pull is defeated by the outward force of angular momentum and forms a layer which is called CENtrifugal force driven BOundary Layer(CENBOL). For a Keplerian orbit,  the centrifugal force is balanced by the gravitational force but once the system turns to be a transonic one, the infalling matter will be piled up at the CENBOL and may create shock waves. Shock waves formation from CENBOL is allowed for a large region of parametric space. We can shrink the parameter space by adding viscosity. But viscosity is such a property that rises the temparature and again the temparature causes the increment in viscosity. So understanding of viscosity in direct way creates a loop.

Initial models of accretion with formation of a disc around BH was made by Prendergust \cite{Burbridge}, Gorbatsky \cite{Gorbatsky} and Shakura \cite{Shakura1972}. The scenario, however, twisted after Shakura and Sunyaev's works \cite{ShakuraSunyaev1973} as they unified all the concepts already discussed and proposed a mechanism of angular momentum transport.
The efficiency of this said mechanism can be parametrized by(later popularly known as Shakura Sunyaev Parameter(SSP)),
\begin{equation}
\alpha_{ss}=\frac{v_t}{c_s} + \frac{H^2}{4\pi \rho c_s^2},
\end{equation}
where $\frac{\rho c_s^2}{2}=\epsilon_r + \frac{3}{2} \rho \frac{k_B T}{m_p}$ is thermal energy density of matter, $\epsilon_r$ is energy density of the radiation, $c_s$ and $v_t$ are the speed of sound inside the fluid and turbulent velocity respectively. $k_B$, $T$, $m_p$, $\rho$ and $H$ are Boltzmann constant, temparature, mass of proton, density and disk height respectively. 
This idea worked quite well in several areas, where the value of the SSP, $\alpha_{ss}$, lies between $0.1$ to $0.4$. Although it was unable to predict some global changes of disc structure, eg.: irregularities in observed $Fe$, $K\alpha$ spectrum of Active Galactic Nuclei. We are to replace the shear velocity $- \omega_{r \phi}$ by the value equal to $\alpha_{ss} \rho c_s^2$.

Studies of accretion procedures with SSP changed the whole pathway. We can find several works. Study of viscous accretion phenomena around rotating gravitational objects with hard surfaces, i.e., mostly around objects like neutron stars and strange objects have done in \cite{viscocitybmukherjee1}. Viscosity due to turbulence, governed by magneto-rotational instability, especially when temparature $T \geq 10^5 K$ is studied \cite{viscocitybmukherjee2}. It has shown that the values of $v_t$ and $\alpha_{ss}$ increase quite rapidly as the disk becomes thinner and thinner. The same trend of using $\alpha_{ss}$ as the representative of viscosity can be seen in different articles \cite{viscocitybmukherjee4,viscosity,viscocitybmukherjee5}. Even there is another interesting fact. There exists a theoretical lower bound of the ratio of the shear viscosity to entropy density (according to the string theory and gauge/ gravity duality), mathematically saying $\frac{\eta}{S} \scriptstyle\gtrsim \frac{\mu \hbar}{4 \pi k_B}$, where $\hbar$ is reduced Planck constants and $\mu \leq 1$. Now say for water this ratio is much higher than the prescribed lower limit. We find some contributions in literature where the author of \cite{viscocitybmukherjee3} has tried to reach near the lower limit considering the accretion process towards a BH to be the background. During this work the shear viscosity part is represented by the help of $\alpha_{ss}$ and it simplified the challenge a lot.
 
After stating a brief hereabouts of the local astrophysical phenomena like accretion we will shift our focus towards the great changes in cosmological studies awkward in late $1990$'s. To justify the late time cosmic accelaration detected by the Ia supernova observations\cite{Perlmutter}, it is required to modify either the 'gravity part' or the 'matter/stress energy part' of the Einstein's field equation. The latter initiated the idea of modelling for Dark Energy(DE) which is a kind of exotic fluid, homoganeously filling the universe. This exerts negative pressure and violates the strong energy condition ($\rho +3 p > 0$) and weak energy condition ($\rho + p > 0$) one by one as the universe keeps on expanding. The best fit DE model ever proposed till date is the cosmological constant $\Lambda=-1$ \cite{Padmanabhancosmologicalconstant}. Except this, there are different proposed candidates of DE. Since 2004, different authors have studied the properties of DE accretion towards a BH. First ever try among them  was done by \cite{Babichev2004} where authors have calculated like F. Michel's paper\cite{Michel} in addition with the Einstein's field equation derived for FLRW metric and have shown that the mass of the BH can be reduced by accretion of phantom energy. Once the expanding universe crosses the phantom barrier and weak energy condition is violated ($\rho + p < 0$) their predictions about the causes of such mass reduction incidents were the accretion of the particles of phantom scalar field. They have compared this with the negetively energised particle creation in the Hawking radiation process. However they have ignored about the back reactions of phantom matter on the BH metric. If the background matter density is low , this ignorance does not make any change. But once the density is comparable to the BH density ~$\frac{M}{R^3} \sim \frac{M}{M^3} \sim \frac{1}{M^2}$, the metric of the BH should be modified significantly. This matter has been studied by\cite{Massdecreasebyphantomenergyaccretion} and they have also shown that the horizon will shrink, but the singularity never become naked. Finite and infinite cosmic time, whenever it may be, the energy density of a cosmological BH will turn zero. These articles are quite sensitive to find the fate of the BH at the future cosmological singularity coined as the `Big Rip'\cite{Bigrip}. Both of these authors have predicted that even long time before the big rip to occur, the BHs will get evaporated away. Now cyclic universe is another  scenario where the cosmos fluctuates but the BHs are to radially increase only. Analysis of \cite{phantomenergyaccretion} indicates that although through phantom energy accretion BHs do not disappear before cyclic universal cosmological turn around, they do not cause problems. A recent day study \cite{10} says the process of phantom accelaration is not as general as suggested by \cite{Babichev2004}, rather the distribution of the scalar field remaining outside the BH would be an interesting area for investigation. In literature we may find many other existing papers where DE accretion on a compact object is studied \cite{Darkenergyaccretion1,Darkenergyacrretion2,Darkenergyacrretion3}.

Now we will look onto galactic rotational curve where the missing mass problem is solved by introducing the idea of some exotic matter called Dark Matter(DM) which is silently present in the galaxy's outer layers and responsible for higher angular momentum observed there. The DM density profile at the core areas of the galaxies is critical to indirect searches but remains poorly constrained. In objects such as M87, the DM profile may be significantly enhanced on subspace scales by the central super massive BHs \cite{darkmatteringalaxycore1}. Currently we can find some papers where the presence of DM  particularly even near to the core of the milky way has been studied \cite{darkmatteringalaxycore2}. An well accepted theory of galaxy formation must account for the large amount of non DM which apparently provides $\geq 80\%$ of the virial mass in clusters like coma and which may constitute massive halos around large galaxies. It has been known for over a decade that tiny gas rich dwarfs favour central DM cores over cusps \cite{dwarfs1,dwarfs2}. Silk and Bloman \cite{SilkandBloman} pointed out that the $cos-B$ sattelite observations of the cosmic flux of $\gamma$-rays can place severe constraints on the density of the DM near the center of the galaxies. Although the precise form of the constraint depends upon which kind of interacting particles are considered (say photinos, higgsinos, shentrinos and neutrinos \cite{Goldberg} with mass m in the GeV range \cite{Lee}). It appears that the data are sensitive to $\gamma$-ray signal from a lump of weakly interacting DM with density of order $10^{-22}$ gm/cc over a region of size $100$-$200$ pc from galactic center. The typical DM density within a radius of order $10^{-24}$ gm/cc. However, the DM density at the center of the galaxy is expected to be much enhanced due to the gravitational attraction of the central component of the galaxy. The central component having mass of order $0.7\times 10^{10}$ $M_{\odot}$, where $M_{\odot}$ is the mass of the sun, then the DM density may reach $4 \times 10^{-21}$ gm/cc.

Several works have been done towards unification of DE and DM. Based on the modelling of the speed of sound as a function of the parameter of EoS \cite{Caplar}, via superfluid Chaplygin gas modelling \cite{Popov}, consideration from the dynamics of a generalised Born-Infield theory \cite{Bento} and many other contributions in literature may be found where the unification or interactions of DM and DE is studied. Very particularly Modified Chaplygin Gas(MCG) is found to be the type of DE representative which efficiently fits when the unification or interactions are to be done. MCG is a DE model which suits enough well for different data. For different values of parameters it shows radiation to phantom. The EoS is given by \cite{Eos1,Eos2,Eos3}
\begin{equation}\label{EoS}
p=\alpha \rho -\frac{\beta}{\rho^n}
\end{equation}
where $\alpha$, $\beta$ and n are different constants. This is why while DE accretion towards a BH is to be considered the MCG is one of the best model to be considered. Previously, Biswas, R. et. al. \cite{Biswasaccretion1} has studied the accretion of MCG on BH and found that the wind is dominating than the accretion and they have commented that the BH accretion disc may get fainter by this process and will ultimately weaken the BH feeding up procedure. Another work \cite{Biswasaccretion2} depicted about the change in flow density through accretion and wind branch. This work even pointed towards a formation of CENBOL.

Many authors have tried to constrain the MCG parameter with different observational data available. We will recall particularly two among them.
First one \cite{dataset1} used type Ia supernovae and BAO data set to predict the best fit parameter values. The values are :\\
\begin{table}[h!]\label{dataset1}
\begin{center}
\begin{tabular}{|>{\centering\arraybackslash}m{3cm}|>{\centering\arraybackslash}m{5cm}|>{\centering\arraybackslash}m{5cm}| }
\hline
Parameter & Best fit values for Constitution+CMB+BAO & Best fit values for Union2+CMB+BAO\\
\hline
$\alpha$ & $0.061 \pm 0.079$ & $0.110 \pm 0.097$\\
\hline
n & $0.053 \pm 0.089$ & $0.089 \pm 0.099$\\
\hline
\end{tabular}
\end{center}
\end{table}\\
However, in some other article \cite{dataset2} authors used Union2, SNIa, OHD, CBF, BAO and CMB data to constrain the MCG model and the best fit parameter values with $1 \sigma$ and $2\sigma$ confidence level are:\\
\begin{table}[h!]\label{dataset2}
\begin{center}
\begin{tabular}{|>{\centering\arraybackslash}m{5cm}|>{\centering\arraybackslash}m{8cm}| }
\hline
Parameter & Best fit values \\
\hline
$\alpha$ & $0.00189^{+0.00583+0.00660}_{-0.00756-0.00915}$ \\
\hline
n & $0.1079^{+0.3397+0.4678}_{-0.2539-0.0.2911}$ \\
\hline
\end{tabular}
\end{center}
\end{table} 

In the present work we wish to incorporate the viscosity. With MCG EoS, the viscous accretion towards a BH will be studied. We will roughly take $\alpha$ very small $\sim 0.05$ and n to be nearly equal to $0.1$. Along with this we will compare the whole system with adiabatic fluid case where we will take the EoS to be $P=K \rho^{\Gamma}$, where the value of $\Gamma$ will in general taken as $1.6$.

In the next section we will construct the mathematics of the model i.e., the construction of first order ODE along with initial values. In the next section to that, we will solve the system numerically and will analyse thoroughly. A brief summary and conclusion will be followed in the final section.
\section{Mathematical Construction of the Model}\label{2}
Bondi formulated spherical accretion and wind for a non-rotating star in Newtonian approach \cite{Bondi}. Here we assume steady and axisymmetric disc. In the article \cite{Biswasaccretion1} authors described accretion and wind flow for inviscid fluid around a compact object, pseudo-Newtonian approach. We here try to study the accretion and the wind flow with viscous fluid parametrized by the parameter $\alpha_{ss}$. The general disc outflow equations are mainly based on well known Navier Stokes equation, since treated the system as fluid flow.
\begin{equation}\label{navierstokes}
\frac{\partial \vec{V}}{\partial t} +\left(\vec{V}.\vec{\bigtriangledown}\right)\vec{V}=\vec{F}-\frac{1}{\rho} \vec{\bigtriangledown}p + \gamma \bigtriangledown^2 \vec{V},
\end{equation}
where $\vec{V}$= velocity vector, $t$= classical time, $\vec{\bigtriangledown} $= classical divergence opertaor, $p$= pressure of the fluid, $\vec{F}$= force vector, $\rho$= density of the fluid, $\gamma$= angular momentum and $\bigtriangledown^2$= classical Laplacian operator. General relativistic equation are highly non-linear and difficult to solve. To simplify we assume the system is in steady state, $\frac{\partial \vec{V}}{\partial t}=0$, but to keep the force factor equivalent to the general relativistic one we will replace it by a pseudo Newtonian force and the corresponding components we get :\\
(a) Radial momentum balance equation:
\begin{equation}\label{radial momentum balance}
u\frac{du}{dx} + \frac{1}{\rho} \frac{dp}{dx}-\frac{\lambda^2}{x^3}+F_g\left( x\right)=0,
\end{equation}
where all the variables are expressed in dimensionless units $u=u_r=\frac{v}{c}$, $x=\frac{r}{r_g}$, $r_g=\frac{GM}{c^2}$, $\lambda$= s where $M$ and $c$ are the mass of the BH and speed of light respectively. $r$, $v$ and $\lambda$ are radial co-ordinate and radial velocity and angular momentum of the disc, $p$ and $\rho$ are dimensionless isotropic pressure and density.
\begin{equation}
F_g\left(x\right)= \frac{\left(x^2-2j\sqrt{x}+ j^2\right)^2}{x^3\lbrace \sqrt{x} \left(x-2\right)+j\rbrace ^2}
\end{equation}
is the gravitational force corresponding to the pseudo Newtonian potential \cite{pseudonewtonian}, where $j$ is dimensionless specific angular momentum for rotating BH.\\
(b) Azimuthal momentum balance equation:
\begin{equation}\label{azimuthal momentum balance}
u \frac{d\lambda}{dx} = \frac{1}{x \Sigma} \frac{d}{dx} \left[ x^2 \alpha_s \left(p+\rho u^2\right) h\left(x\right) \right],
\end{equation}
where $\Sigma$ is vertically integrated density given by,
\begin{equation}
\Sigma = I_c \rho_e h\left(x\right),
\end{equation}
when $I_c$= constant (related to equation of state of fluid)= $1$ \cite{Biswasaccretion1}, 
$\rho_e$=density at equitorial plane, 
$h\left(x\right)$= half thickness of the disc.\\
Assuming the vertical equilibrium from the vertical component of equation (\ref{navierstokes}) we get the expression for $h(x)$ as
\begin{equation}\label{halfthikness}
h\left(x\right) = c_s \sqrt{\frac{x}{F_g}}.
\end{equation}
(c) The vertically integrated mass conservation relation i.e., the equation of continuity for disc accretion is 
\begin{equation}\label{equation of continuity}
\frac{d}{dx} \left(x u \Sigma\right) = 0.
\end{equation}
Differentiating equation (\ref{EoS}) with respect to $\rho$ we have,
\begin{equation}\label{Eoq2}
c_s^2 = \frac{\partial p}{\partial \rho} = \alpha + \frac{\beta n}{\rho^{n+1}}.
\end{equation}
Thus we get from the above equations (\ref{equation of continuity}) and (\ref{Eoq2})\begin{equation}\label{Eoq3}
\frac{1}{\rho} \frac{dp}{dx} = - \frac{2 c_s^3}{\left(n+1\right) \left( c_s^2 - \alpha \right)} \frac{d c_s}{dx} =-\frac{1}{n+1} \frac{d}{dx} \left(c_s^2 \right) - \frac{\alpha}{n+1} \frac{d}{dx} \lbrace ln \left( c_s^2 -\alpha \right) \rbrace.
\end{equation}
Now integrating the equation (\ref{equation of continuity}) we get the mass conservation equation
\begin{equation}\label{mass conservation}
\dot{M} = \Theta \rho c_s \frac{x^\frac{3}{2}}{F_g^\frac{1}{2}} u,
\end{equation}
where $\Theta$ is geometrical constant.\\
Replacing the value of $\rho$ from equation (\ref{Eoq2}) in (\ref{mass conservation}) and differentiating the whole term we get a differential equation for $c_s$,
\begin{equation}\label{differential equation for c}
\frac{d c_s}{dx} = \left( \frac{3}{2x} -\frac{1}{2F_g} \frac{dF_g}{dx} + \frac{1}{u} \frac{du}{dx} \right) \left\lbrace \frac{\left( n+1 \right) c_s \left( c_s^2 -\alpha \right)}{\left( 1-n \right) c_s^2 + \alpha \left( n+1 \right)} \right\rbrace 
\end{equation}
and from the equations (\ref{azimuthal momentum balance}) and (\ref{EoS}) we get,
\begin{multline}\label{differential equation for lambda}
\frac{d \lambda}{dx} = \frac{x \alpha_s}{u} \left[ \frac{1}{2} \left( \frac{5}{x} - \frac{1}{F_g} \frac{dF_g}{dx} \right) \left\lbrace \frac{\left(n+1 \right) \alpha - c_s^2}{n} + u^2 \right\rbrace \right.\\
\left. + 2 u \frac{du}{dx} + \left\lbrace \left( \frac{\left(n+1 \right) \alpha - c_s^2}{n} + u^2 \right)  \frac{1}{c_s} -\left( c_s^2 +u^2 \right) \left( \frac{1}{n+1} \frac{2 c_s}{c_s^2 - \alpha} \right) \right\rbrace \frac{dc_s}{dx} \right]. 
\end{multline}
Thus we get all the needed values to replace $\frac{1}{\rho} \frac{dp}{dx}$ in the equation (\ref{radial momentum balance}), hence we obtain
\begin{equation}\label{differential equation for u}
\frac{du}{dx} = \frac{\frac{\lambda^2}{x^3}- F_g\left( x \right) + \left( \frac{3}{x} - \frac{1}{F_g} \frac{dF_g}{dx} \right) \frac{ c_s^4}{\lbrace \left( 1-n \right) c_s^2 + \alpha \left( n+1 \right)\rbrace}}{u - \frac{2 c_s^4}{u \lbrace \left( 1-n \right) c_s^2 + \alpha \left( n+1 \right)\rbrace}}.
\end{equation}
As unlike a nutron star BH does not posses any hard surface rather than there is a event horizon, the flow towards a BH is necessarily transonic. Far from the BH the accretion speed must be very low and near to the horizon this should be equal to the speed of light. So it is clear that in between somewhere the fluid flow velocity must be equal to the speed of sound inside it. This point is called sonic point. We can observe the denominator of (\ref{differential equation for u}) will be zero at a particular value of $u$ and $c_s$.Therefore for the sake of the stability of the disc the numerator also have to zero. This point of the disc is called critical point. Now applying the L'hospital rule  in equation (\ref{differential equation for u})at the critical point (say $x =x_c$) we get a quadratic equation of $\frac{du}{dx}$ in the form,
\begin{equation}\label{quadratic}
A \left( {\frac{du}{dx}}\right)_{x=x_c}^2 + B \left( \frac{du}{dx} \right)_{x=x_c} + C =0.
\end{equation}
Where \\
$A= 1 + \frac{1}{c_{sc}^2} - \frac{4 \left( n+1 \right) \left( c_{sc}^2 -\alpha \right)}{{\left\lbrace \left( 1-n \right) c_{sc}^2 +\alpha \left( n+1 \right) \right\rbrace}^2}$,\\
$B= \frac{4 \lambda u}{x_c^3} + \frac{2 u_c c_{sc} \left( 1-n \right)}{ \left\lbrace \left( 1-n \right) c_{sc}^2 +\alpha \left( n+1 \right) \right\rbrace} - \frac{2 u_c \left( n+1 \right) \left( c_{sc}^2 -\alpha \right)}{{\left\lbrace \left( 1-n \right) c_{sc}^2 +\alpha \left( n+1 \right) \right\rbrace}^2} \left( \frac{3}{x_c} - \frac{1}{F_g} \left(\frac{dF_g}{dx}\right)_{x=x_c} \right) + \\
\left[ \left( c_{sc}^2 +u_c^2 \right) \left( \frac{1}{n+1} \frac{2 c_{sc}}{c_{sc}^2 - \alpha} \right)-\left( \frac{3}{x_c} - \frac{1}{F_g} \left(\frac{dF_g}{dx}\right)_{x=x_c} \right) \frac{4 u_c }{c_{sc}{\left\lbrace \left( 1-n \right) c_{sc}^2 +\alpha \left( n+1 \right) \right\rbrace}}-  \left( \frac{\left(n+1 \right) \alpha - c_{sc}^2}{n} + u^2 \right)  \frac{1}{c_{sc}} \right] \left\lbrace \frac{ u_c \left( n+1 \right) \left( c_{sc}^2 -\alpha \right)}{2 c_{sc}^3} \right\rbrace $\\
and $C= D+ E+ F$.\\
The values of $D$, $E$ and $F$ are,\\
$D= \left[ \left( c_{sc}^2 +u_c^2 \right) \left( \frac{1}{n+1} \frac{2 c_{sc}}{c_{sc}^2 - \alpha} \right)-\left( \frac{3}{x_c} - \frac{1}{F_g} \left(\frac{dF_g}{dx}\right)_{x=x_c} \right) \frac{4 u_c }{c_{sc}{\left\lbrace \left( 1-n \right) c_{sc}^2 +\alpha \left( n+1 \right) \right\rbrace}}-  \left( \frac{\left(n+1 \right) \alpha - c_{sc}^2}{n} + u^2 \right)  \frac{1}{c_{sc}} \right] \\
\left( \frac{3}{2x_c} -\frac{1}{2F_g} \left(\frac{dF_g}{dx} \right)_{x=x_c} \right) \left\lbrace \frac{\left( n+1 \right) c_{sc} \left( c_{sc}^2 -\alpha \right)}{\left( 1-n \right) c_{sc}^2 + \alpha \left( n+1 \right)} \right\rbrace $,\\
$E= \left( \frac{3}{x_c} - \frac{1}{F_g} \left(\frac{dF_g}{dx}\right)_{x=x_c} \right) \frac{u_c c_{sc} \left( 1-n \right)}{{\left\lbrace \left( 1-n \right) c_{sc}^2 +\alpha \left( n+1 \right) \right\rbrace}} + \left(\frac{dF_g}{dx}\right)_{x=x_c} + \left\lbrace \frac{1}{F_g^2} \left(\frac{dF_g}{dx} \right)^2_{x=x_c} - \frac{1}{F_g} \left(\frac{d^2 F_g}{dx^2}\right)_{x=x_c} - \frac{3}{x_c^2}\right\rbrace \frac{u_c^2}{2}$,\\
$F= \frac{\lambda \alpha_{ss}}{x_c^2 u_c} \left( \frac{1}{F_g} \left(\frac{dF_g}{dx}\right)_{x=x_c} - \frac{5}{x_c} \right) \left\lbrace \frac{\left(n+1 \right) \alpha - c_{sc}^2}{n} + u_c^2 \right\rbrace  $.

While we are working with a particular type of EoS we supply $n$ and $\alpha$, i.e., vanishing denominator, $D\left( u_c, c_{sc}, n, \alpha\right)=0$ will be reduced to an equation $D\left( u_c, c_{sc}\right)=0$, which tell us the relation between $u_c$ and $c_{sc}$, on the other hand vanishing numerator, i.e.,$N\left( \lambda_c, x_c, j, c_{sc}, n, \alpha\right)=0$ will be reduced to $N\left( \lambda_c, x_c, c_{sc}\right)=0$ once we provide the information about the accreting fluid and the rotation of the BH. We will choose where the sonic point will be formed and adjust the value of angular momentum at the critical point which ultimately form an algebraic equation of $c_{sc}$,  $N \left( c_{sc}=0 \right)$ From there we can easily solve $c_{sc}$ and hence calculate the value of $u_c$. These will be used as the initial values to solve (\ref{quadratic}). As we proceed towards the BH from the critical point we can have two solutions due to the quadratic nature of (\ref{quadratic}). Same will happen if we receed far from the BH. From this new four points (\ref{differential equation for c}), (\ref{differential equation for lambda}), (\ref{differential equation for u}) we can build the whole solution sets. In the next section we will plot the solutions and analyse them physically.             

\section{Numerical Solutions and Their Physical Interpretations}

We will plot the numerically derived solutions in this section. Figure $1.1.a$ and $1.1.b$ are the plots of $log(u)$ vs $log(x)$ for adiabatic flow and MCG flow respectively. The solid line is for accretion and the dotted line is showing the behaviour of wind. The set $1.1.a$ to $1.3.b$ is drawn for a nonrotating BH and the viscosity is increased gradually.
\begin{figure}[h!]
\centering
~~~~~~~Fig $1.1.a$~~~~~~~\hspace{3 in}~~~~~~~~~Fig $1.1.b$
\includegraphics*[scale=0.7]{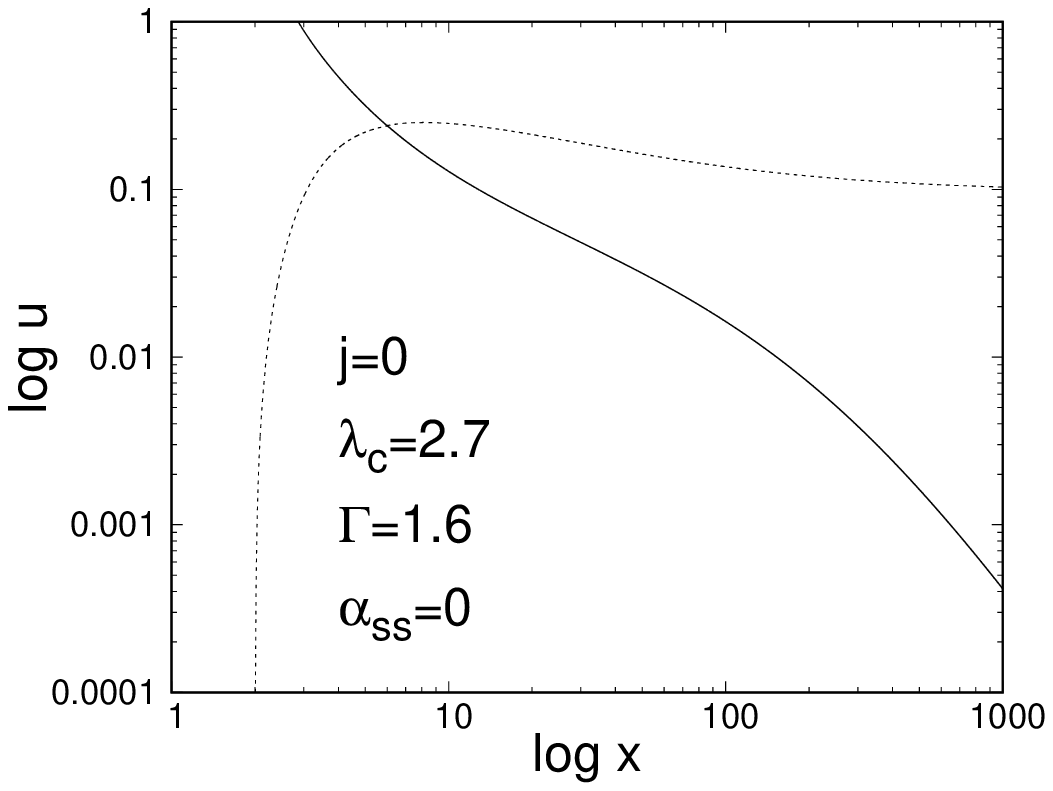}~~
\hspace*{0.2 in}
\includegraphics*[scale=0.7]{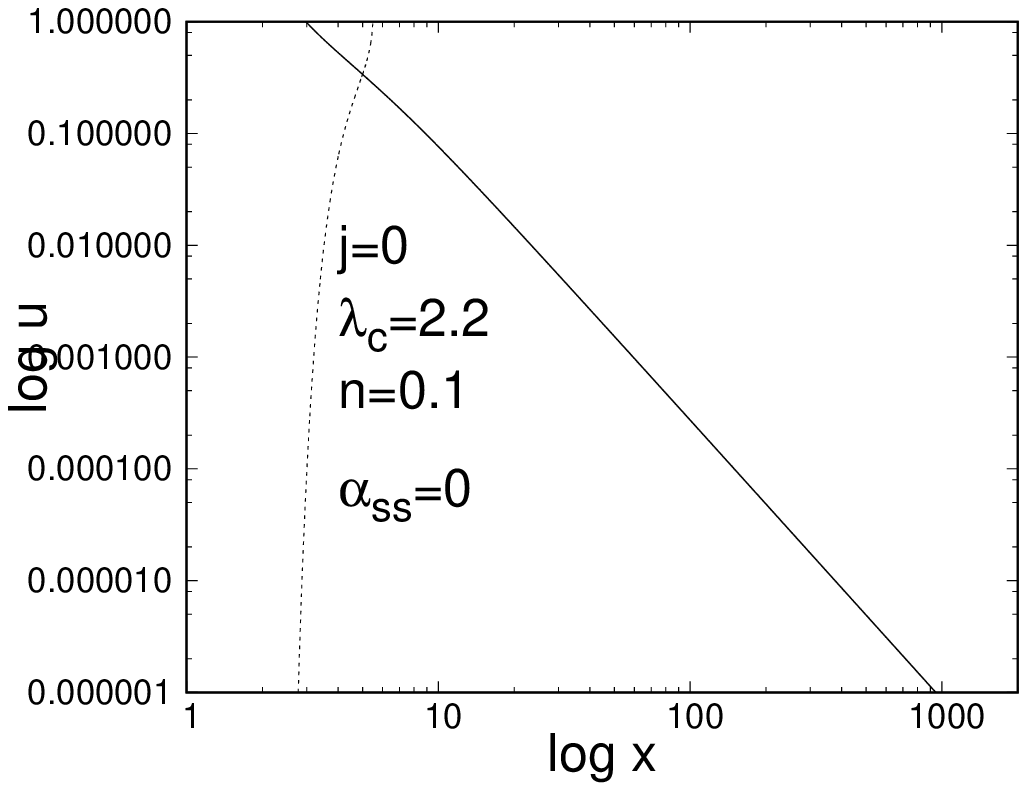}~~\\
\it{Figures $1.1.a$ and $1.1.b$ are plots of $log(u)$ vs $log(x)$ for nonviscous accretion disc flow around a non-rotating BH for adiabatic and MCG respectively. Accretion and wind curves are depicted by solid and dotted lines respectively.}
\end{figure}

A popular explanation towards the angular momentum transpotation is given in such a way that whenever a particle is falling into a BH the momentum lost by it is gained by a farthest point particle, which starts just to rotate, i.e., enters in the disc. We think every layers's rotation is a cause of the rotation of just the adjacent next layer to it. If we go outwards the angular momentum increases but the angular velocity decreases. This causes the increment of the radial inward velocity towards the central engine. When a particle starts up at a distance of thousand Schwarzschild radius, the inward radial velocity is low. This increases gradually and at some point near to the central gravitating object, the flow velocity meets the speed of sound inside the fluid, i.e., reaches the sonic point. Now how far from the BH  the sonic point will be formed is manually chosen for this figure. If we go near to the BH, density $\rho$ is high and the speed of sound is also high. This depicts $\frac{\rho c_s^2}{2}$, i.e., the energy of the accreting fluid should be high. We are setting the sonic point near to the BH means we are giving it high amount of energy. For we set the sonic point at far less the energy is given to it.
After crossing the sonic point speed, as BH accretion is a transonic process, the radial inward speed steeply increase and near to the BH event horizon it is almost equal to the speed of light. In accretion branch at the farthest point, where the radial velocity is almost zero, we can find that the specific angular momentum is nearly equal to the Keplerian angular momentum there. So once we cross the radial distance  and go further which is such strong that the centrifugal force created will defeat the inward attraction. We will analyse the wind branch which is less near to the BH which is obvious due to the strong gravitational force there. This strong inward force will not allowed anything to get throughout. As we increase our radial distance from the BH, this wind speed increases and it reaches a local maximum at a distance $x_{max}^{wind} > x_c$, the sonic point distance. If $x > x_{max}^{wind} $ the graph of velocity gradually decreases and the rate of which is low. This shows even far from a BH we can feel the outward throwing force of it upon us. 

\begin{figure}[h!]
\centering
~~~~~~~Fig $1.2.a$~~~~~~~\hspace{3 in}~~~~~~~~~Fig $1.2.b$
\includegraphics*[scale=0.7]{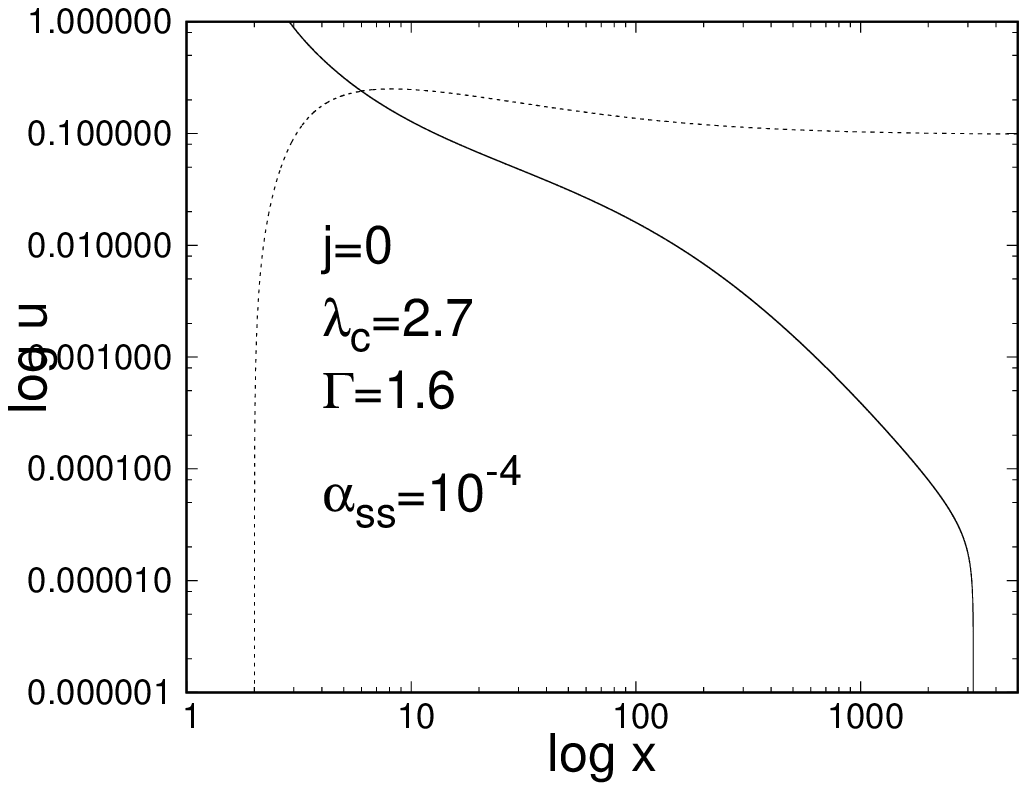}~~
\hspace*{0.2 in}
\includegraphics*[scale=0.7]{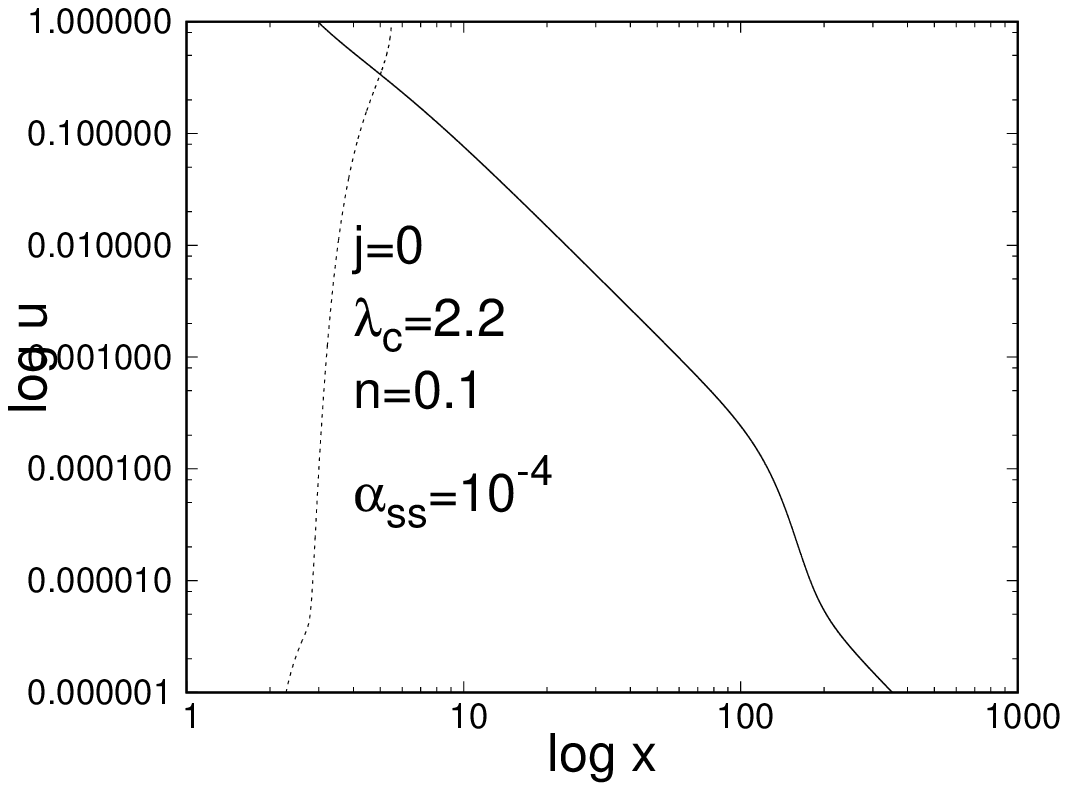}~~\\
\it{Figures $1.2.a$ and $1.2.b$ are plots of $log(u)$ vs $log(x)$ for viscous ($\alpha_{ss}=10^{-4}$) accretion disc flow around a non-rotating BH for adiabatic and MCG respectively. Accretion and wind curves are depicted by solid and dotted lines respectively.}
\end{figure}
This scenario changed a lot when we change the type of the EoS and use the equation of MCG. First, the radial velocity to get started, the fluid has to come nearer to the BH. DE is something which exerts negative pressure and tries not to be bound in some predicted volume and tries to expand. So to bind it in a astrophysical phenomeana like accretion disc, we need to go near to the BH to feel more force of attraction. Once it starts to accrete its inward radial speed increases and become almost equal to the speed of light near the BH. So inner region properties of accretion branch for DE are more or less matching with that of the adiabatic flow. The huge difference we can follow in the wind branch. Wind, unlike the accretion branch, inspite of being almost constant after a distance, becomes almost equal to the speed of light at a finite distance. This says the matter is thrown apart from the BH accretion disc with a speed equal to light there. This again support the nature of the DE which be exerting negative pressure enriches the outwards motion.
Upto this we have rebuilt and enriched the result found by the work \cite{Biswasaccretion1}.

\begin{figure}[h!]
\centering
~~~~~~~Fig $1.3.a$~~~~~~~\hspace{3 in}~~~~~~~~~Fig $1.3.b$
\includegraphics*[scale=0.7]{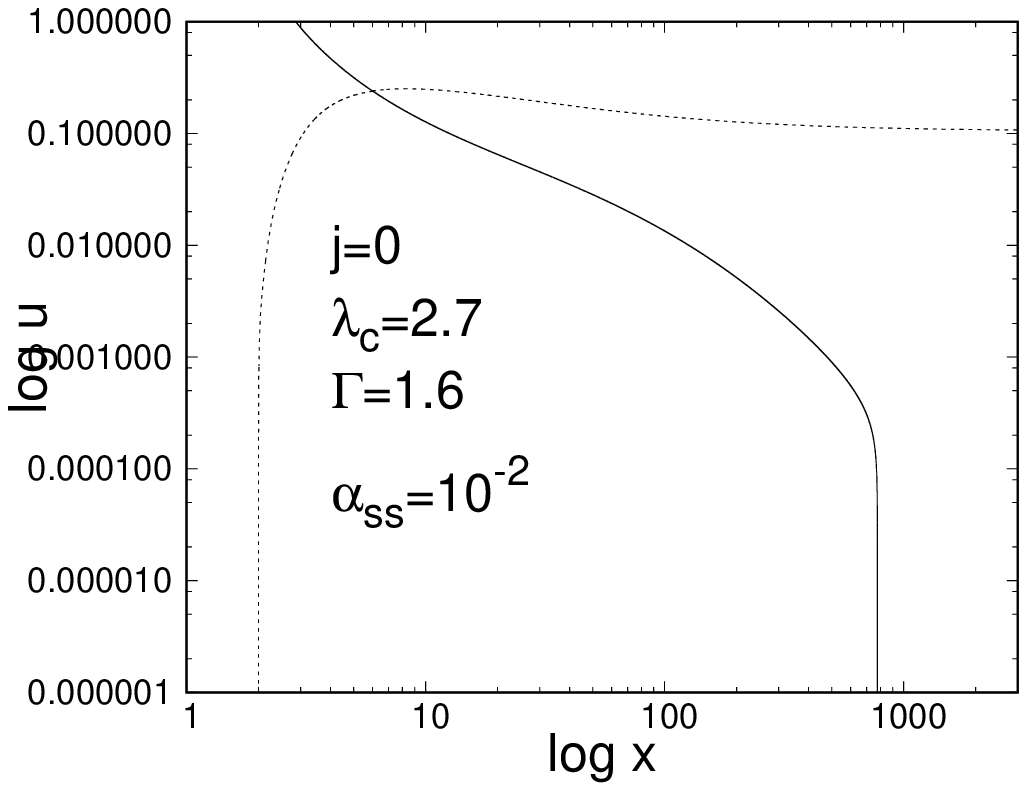}~~
\hspace*{0.2 in}
\includegraphics*[scale=0.7]{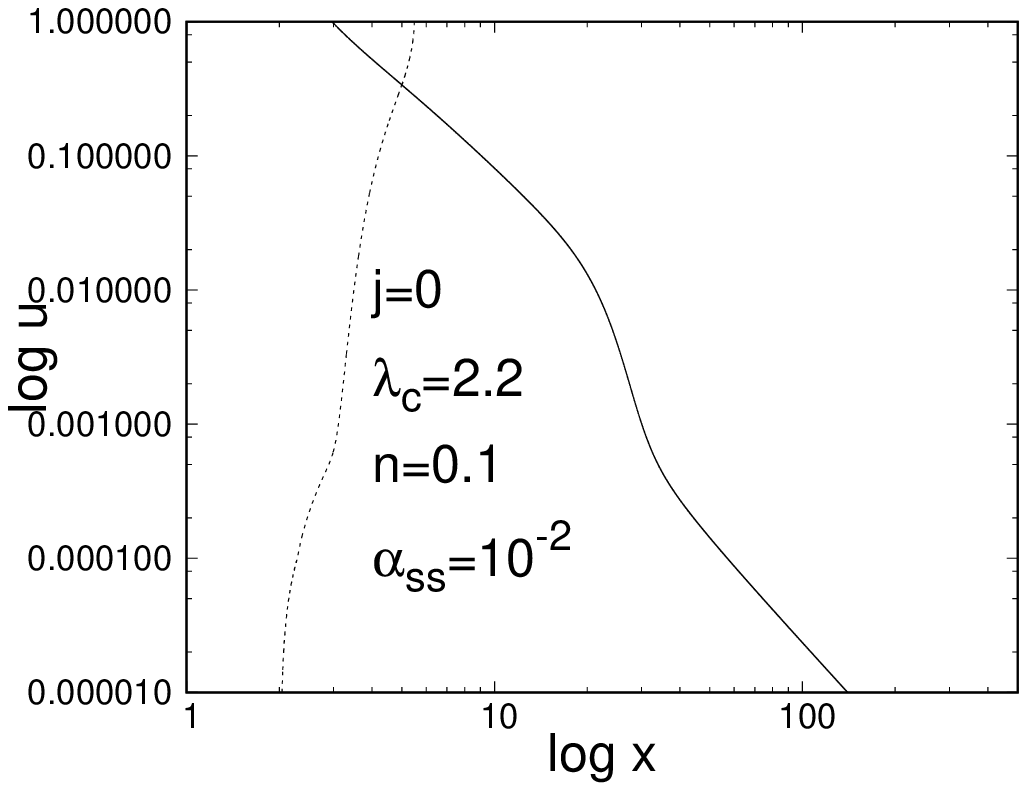}~~\\
\it Figures $1.3.a$ and $1.3.b$ are plots of $log(u)$ vs $log(x)$ for viscous ($\alpha_{ss}=10^{-2}$) accretion disc flow around a non-rotating BH for adiabatic and MCG respectively. Accretion and wind curves are depicted by solid and dotted lines respectively.
\end{figure}

What we have done next, added viscosity to the system. The parameter $\alpha_{ss}$ will regulate the value of the viscosity throughout. First we take very low $\alpha_{ss}=10^{-4}$. We have plotted the viscous accretion cases in the figures $1.2.a$ and $1.2.b$. The first change we can see in the plot of $log(u)$ vs $log(x)$ in Figure $1.2.a$ where the fluid is adiabatic and $j=0$. We see the accretion disc last upto several hundreds of Schwarzschild radius only. Why this shortening? We can explain this in two ways. Firstly, from the outward angular momentum transportation idea we can say as the momentum is transported layer after layer outwards, due to the viscosity some more part of it is transported than the inviscid one and soon it becomes Keplerian. To revise the angular momentum more radial velocity is lost and nearer than the inviscid case we find, the edge of the disc. Secondly, the matter which is coming from outwards whenever tries to get in accretion procedure is opposed by the opposite viscous force/ frictional force between two outer layer coming from the viscosity. This incident forces the accretion particle to come to more nearer region to set the accretion phenomenon to get started. The accretion of MCG is showing the same pattern in the Figure $1.1.b$. Only the difference is the radius of accretion disc is more reduced than an adiabatic one.

\begin{figure}[h!]
\centering
~~~~~~~Fig $2.1.a$~~~~~~~\hspace{3 in}~~~~~~~~~Fig $2.1.b$
\includegraphics*[scale=0.7]{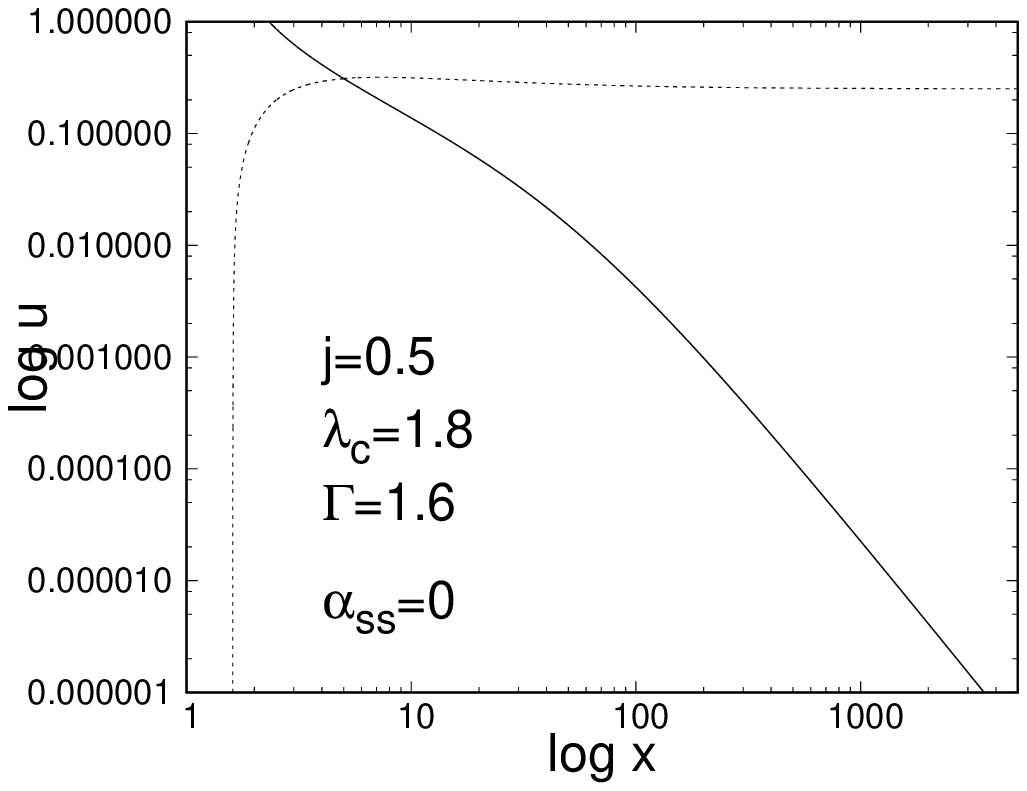}~~
\hspace*{0.2 in}
\includegraphics*[scale=0.7]{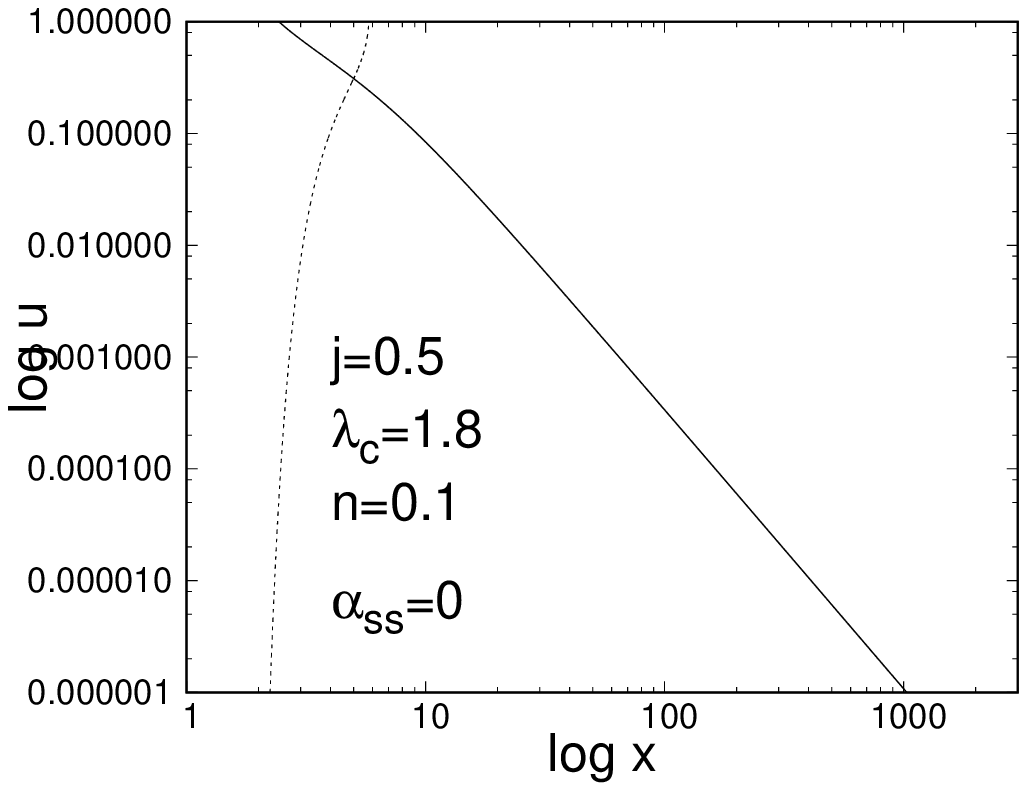}~~\\
\it{Figures $2.1.a$ and $2.1.b$ are plots of $log(u)$ vs $log(x)$ for nonviscous accretion disc flow around a rotating BH (with specific angular velocity $j=0.5$)for adiabatic and MCG respectively. Accretion and wind curves are depicted by solid and dotted lines respectively.}
\end{figure}
We plot $log(u)$ vs $log(x)$ for $\alpha_{ss}=10^{-2}$ in the figures $1.3.a$ and $1.3.b$. We see the accretion disc length to get shortened. This happens only due to high rate outward transportation of angular momentum and after that $\frac{\lambda}{\lambda_k}$ exceeds unity. this mean beyond the distance the disc is super-Keplerian. For the wind branch of the MCG we see as we increase viscosity the inner part of the wind branches able to go near to the BH. On the other hand, adiabatic flow, as we go far from BH drops down rapidly. But for MCG even we go far, we can see a mild amount of accretion continues.

Next set of curves are drawn for the specific angular velocity $j=0.5$. Figures $2.1.a$ and $2.1.b$ are the inviscid cases for adiabatic and MCG respectively. Increment in rotation of BH decrease the gravitating inward force. This is because pseudo Newtonian force is a decreasing function of $j$. This attracts the particle less. So angular momentum is transported more vigorously outward and sooner it reaches the point where $\frac{\lambda}{\lambda_k}$ becomes unity. This is why se see the accretion to stop nearer than the non rotating case. We plot $\alpha_{ss}=10^{-4}$ and $10^{-2}$ cases for BH with rotation $j=0.5$ in figures $2.2.a$ and $2.3.a$ (adiabatic) and figures $2.2.b$ and $2.3.b$ (MCG) respectively. Overall trends follows the trend of figures drawn in set $1.1.a$ to $1.3.b$.
 
\begin{figure}[h!]
\centering
~~~~~~~Fig $2.2.a$~~~~~~~\hspace{3 in}~~~~~~~~~Fig $2.2.b$
\includegraphics*[scale=0.7]{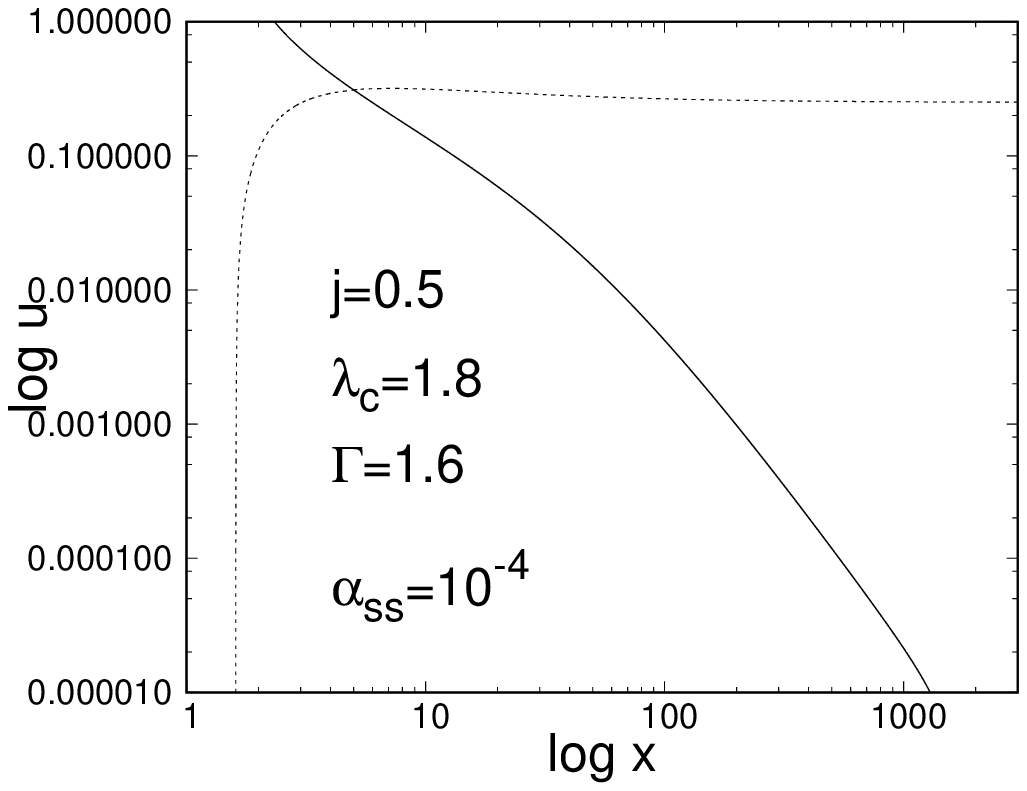}~~
\hspace*{0.2 in}
\includegraphics*[scale=0.7]{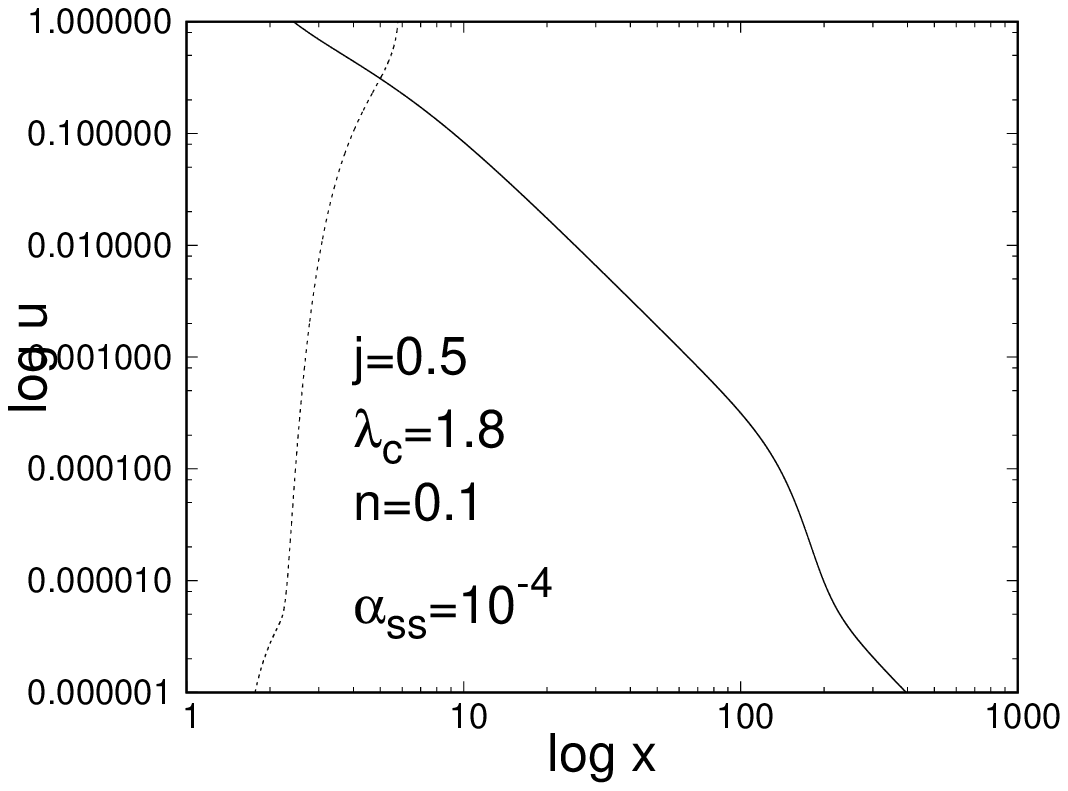}~~\\
\it{Figures $2.2.a$ and $2.2.b$ are plots of $log(u)$ vs $log(x)$ for viscous ($\alpha_{ss}=10^{-4}$) accretion disc flow around a rotating BH (with specific angular velocity $j=0.5$) for adiabatic and MCG respectively. Accretion and wind curves are depicted by solid and dotted lines respectively.}
\end{figure}
\begin{figure}[h!]
\centering
~~~~~~~Fig $2.3.a$~~~~~~~\hspace{3 in}~~~~~~~~~Fig $2.3.b$
\includegraphics*[scale=0.7]{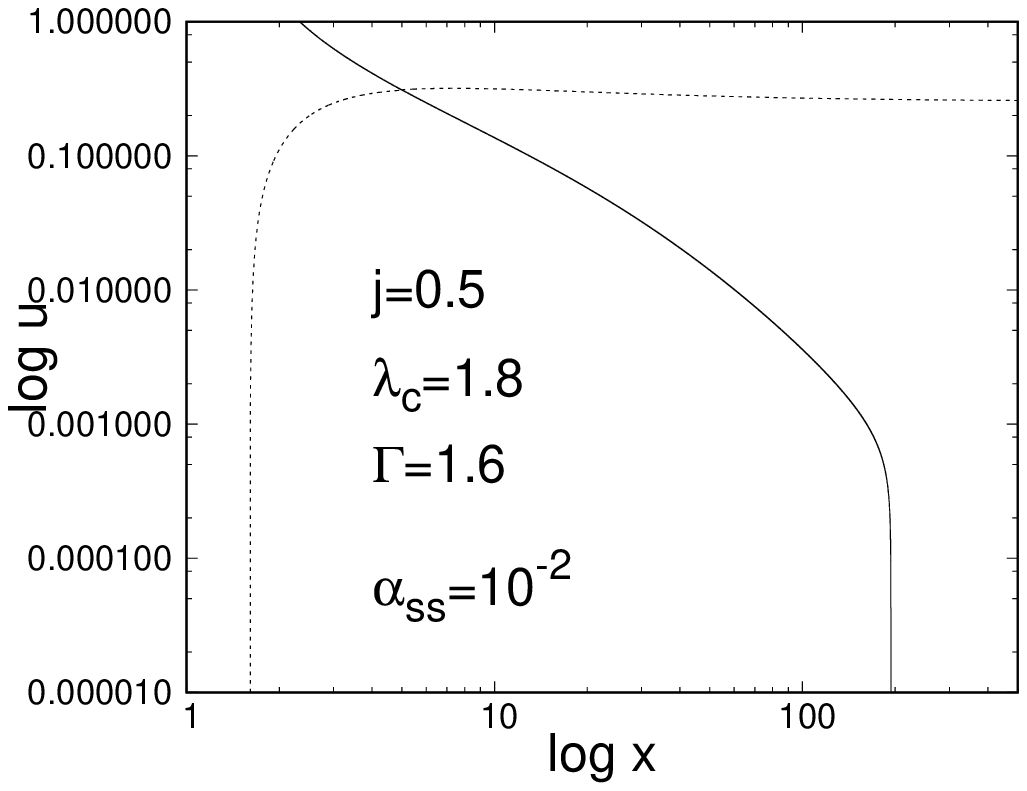}~~
\hspace*{0.2 in}
\includegraphics*[scale=0.7]{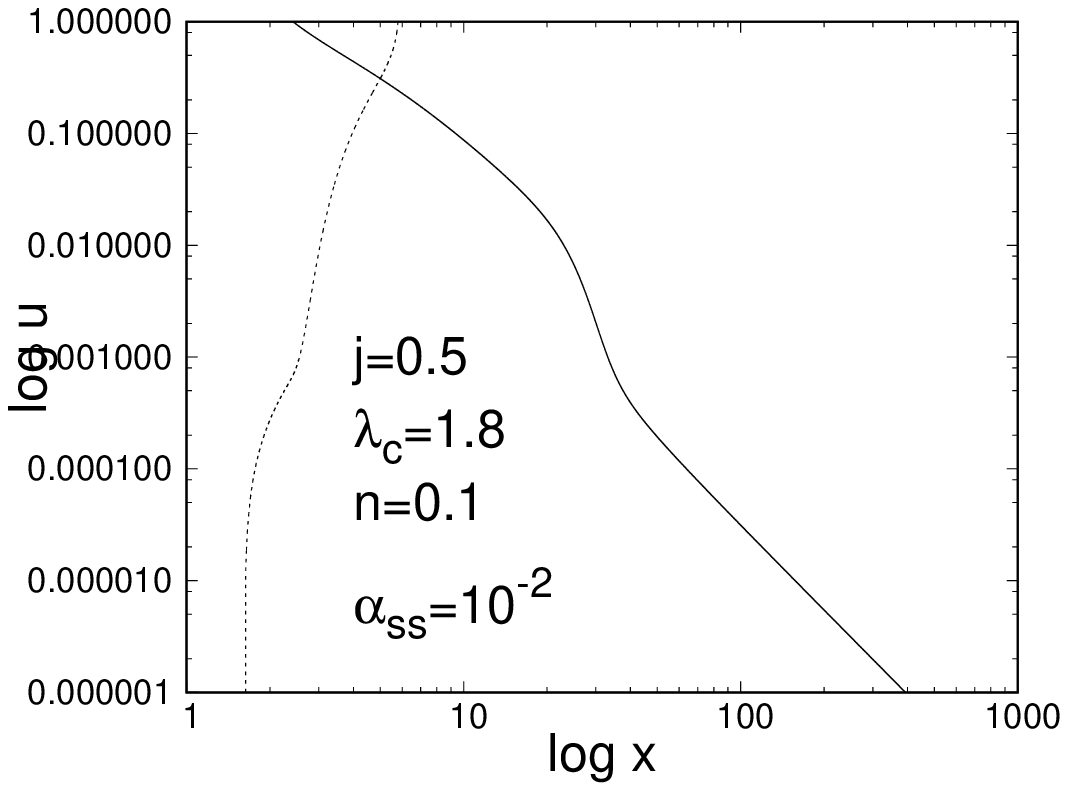}~~\\
\it{Figures $2.3.a$ and $2.3.b$ are plots of $log(u)$ vs $log(x)$ for viscous ($\alpha_{ss}=10^{-2}$) accretion disc flow around a rotating BH (with specific angular velocity $j=0.5$) for adiabatic and MCG respectively. Accretion and wind curves are depicted by solid and dotted lines respectively.}
\end{figure}

We plot figures $3.1.a - 3.3.b$ for $j=0.9$. Keeping the general trends same we see that the accretion disc lenth is reduced as we increase the rotation parameter $j$. As a whole we can conclude that  viscosity shortens the physical length of the accretion disc the same is done by the DE accretion. again if the rotation parameter of the BH is high the disc length reduces more.
\begin{figure}[h!]
\centering
~~~~~~~Fig $3.1.a$~~~~~~~\hspace{3 in}~~~~~~~~~Fig $3.1.b$
\includegraphics*[scale=0.7]{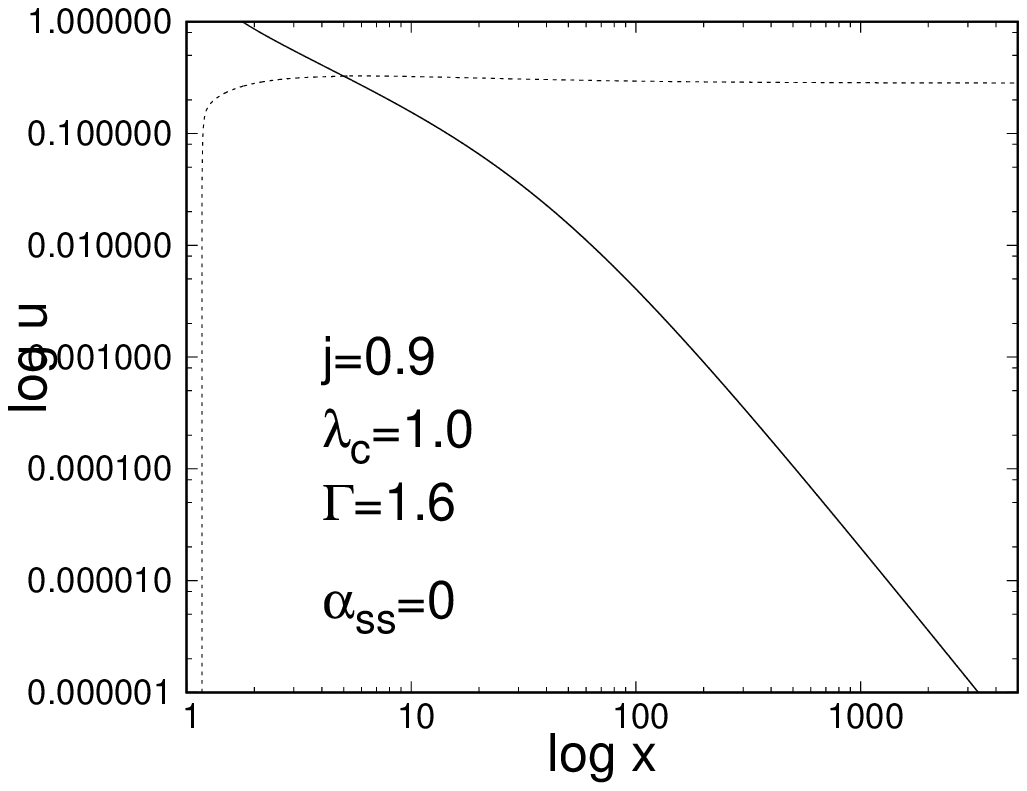}~~
\hspace*{0.2 in}
\includegraphics*[scale=0.7]{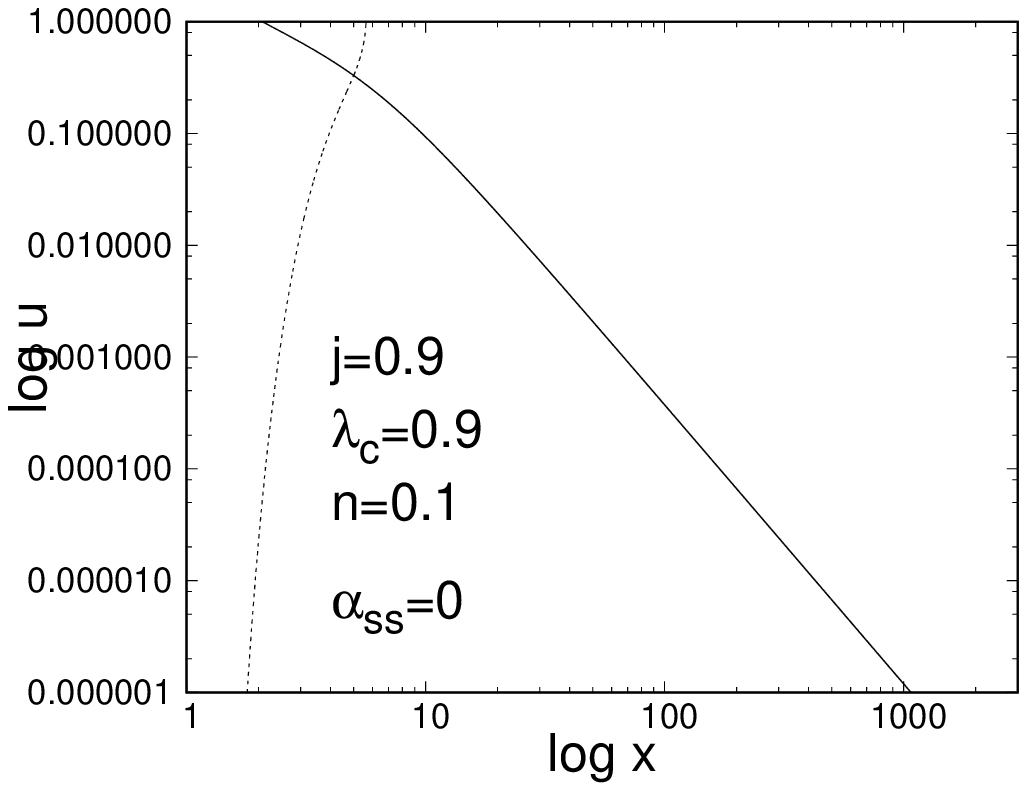}~~\\
\it{Figure $3.1.a$ and $3.1.b$ are plots of $log(u)$ vs $log(x)$ for nonviscous accretion disc flow around a rotating BH (with specific angular velocity $j=0.9$)for adiabatic and MCG respectively. Accretion and wind curves are depicted by solid and dotted lines respectively.}
\end{figure}
\begin{figure}[h!]
\centering
~~~~~~~Fig $3.2.a$~~~~~~~\hspace{3 in}~~~~~~~~~Fig $3.2.b$
\includegraphics*[scale=0.7]{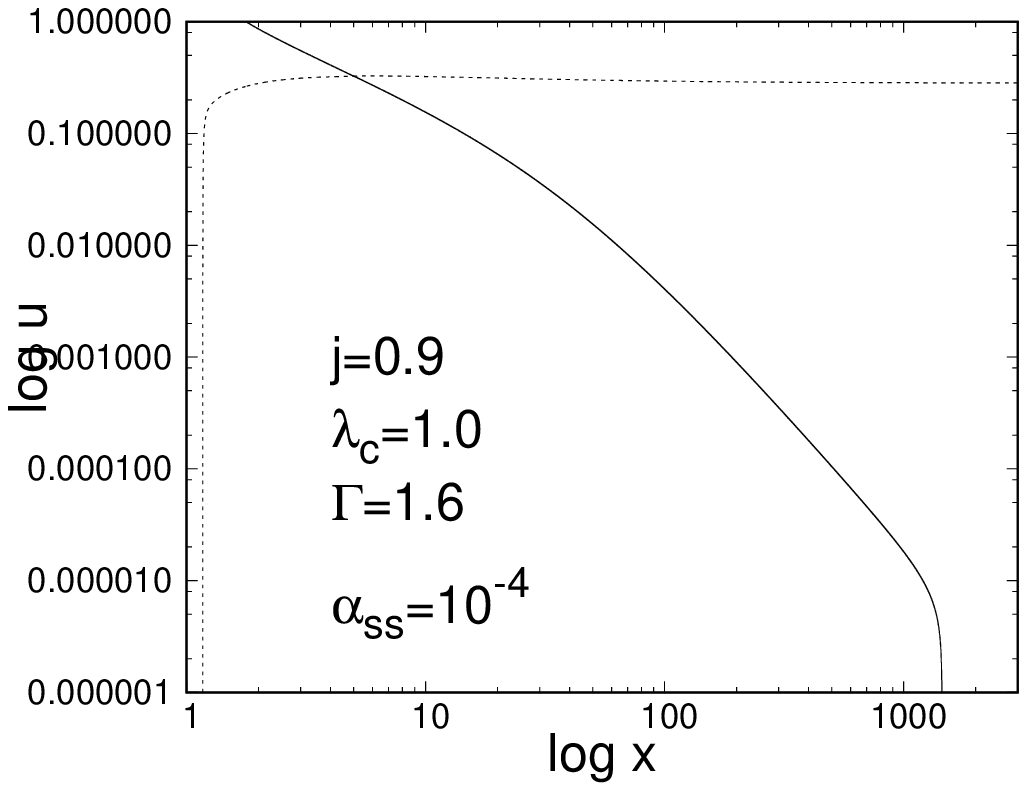}~~
\hspace*{0.2 in}
\includegraphics*[scale=0.7]{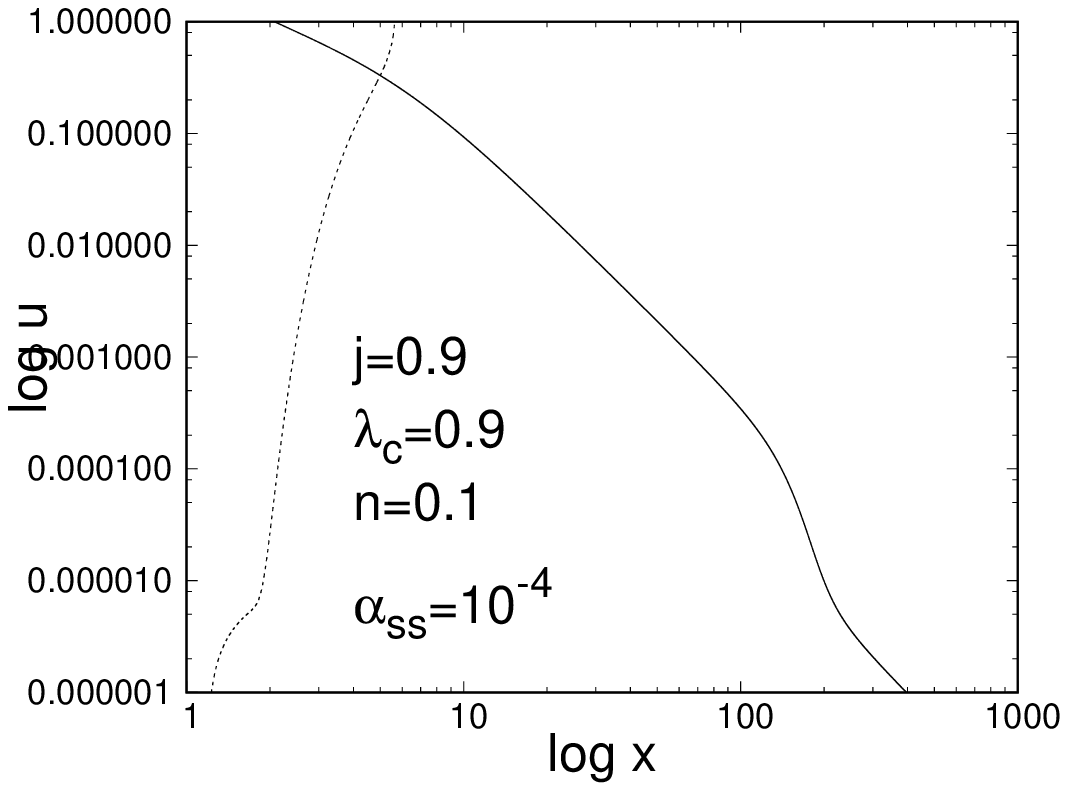}~~\\
\it{Figures $3.2.a$ and $3.2.b$ are plots of $log(u)$ vs $log(x)$ for viscous ($\alpha_{ss}=10^{-4}$) accretion disc flow around a rotating BH (with specific angular velocity $j=0.9$) for adiabatic and MCG respectively. Accretion and wind curves are depicted by solid and dotted lines respectively.}
\end{figure}
\begin{figure}[h!]
\centering
~~~~~~~Fig $3.3.a$~~~~~~~\hspace{3 in}~~~~~~~~~Fig $3.3.b$
\includegraphics*[scale=0.7]{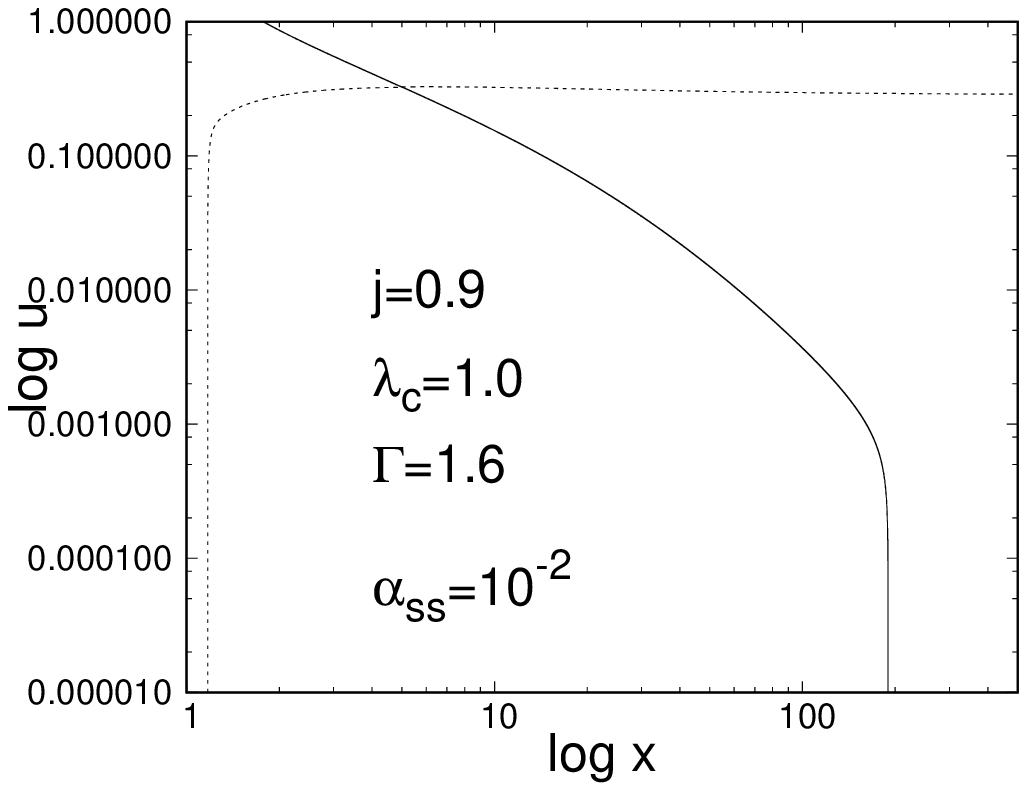}~~
\hspace*{0.2 in}
\includegraphics*[scale=0.7]{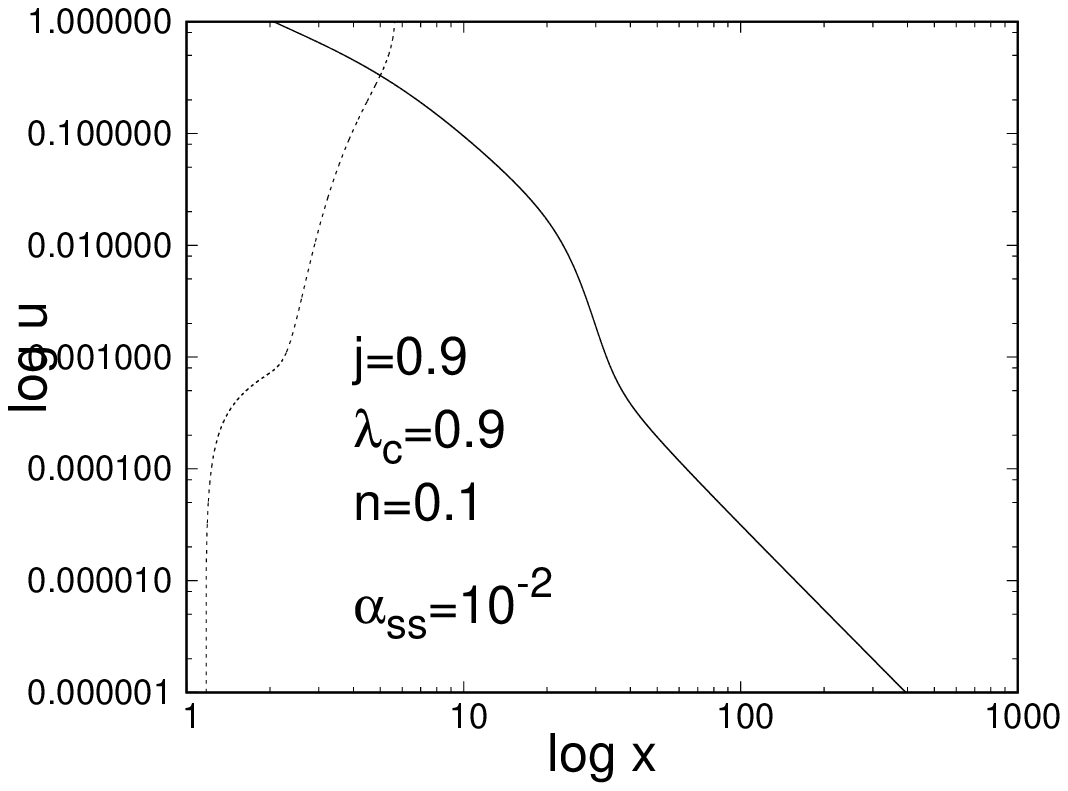}~~\\
\it{Figures $3.3.a$ and $3.3.b$ are plots of $log(u)$ vs $log(x)$ for viscous ($\alpha_{ss}=10^{-2}$) accretion disc flow around a rotating BH (with specific angular velocity $j=0.9$) for adiabatic and MCG respectively. Accretion and wind curves are depicted by solid and dotted lines respectively.}
\end{figure}

\section{Brief Summary and Conclusion}

This paper deals with MCG accretion on BHs. We have studied the viscous and non viscous flows of both adiabatic and MCG. The central engine has been treated to be non-rotating and rotating both. We have incorporated the viscosity via the introduction of SSP. Mainly we have solved the three components of Navier-Stoke's equation along with the equation of continuity and the equation of state numerically. To make the transonic flow continuous throughout the radial distance traversed from very far region to very near to a BH. We have taken two branches - accretion and wind, which will coincide the sonic point. Accretion branches shows the radial inward velocity of accreting matter towards the BH. Wind on the other hand depicts the velocity with which the fluid go far from the central engine. For adiabatic flow we see radial inward velocity increases as we go towards BH and wind decreases towrads BH. For MCG wind increases as as we go far from the BH and becomes equal to the speed of light at a finite distance. This says MCG throws out material from the accretion disc and weakens that. when we include rotation to the central engine, i.e., a BH, we see that for adiabatic cases the behaviour of the viscous accretion flow is same as the non-rotating case untill some distance, then suddenly falls. For MCG accretion with viscosity, while the central BH is rotating, both accretion and wind flow reaches its climax earlier which means the length of the disc shortened. So we can say somehow the specific angular velocity of the central engine helps MCG to weaken the disc.we have also seen that the specific angular momentum of the disc flow effects accretion and wind. The behaviour of accretion and wind flow is on an average same when we increase specific angular momentum upto a limit, but for extremely high value the accretion disc no longer exist for both adiabatic and MCG.

Accretion procedure feeds up the BH and the mass increases due to this feeding up process. If MCG is throwing the accreting material out of the the disc, it weakens the BH indirectly as we increase the viscosity, outward angular momentum transport is more efficient along with the negatively pressure creating fluid, MCG. So viscosity catalyze the effect of MCG. This supports previous works done with DE accretion where in future towards Big-Rip will weaken the BH so much that the BH may not exist at that point. 

\section{Acknowledgement}

Authors thank IUCAA, Pune for local hospitality. A part was done during a visit there. RB thanks IUCAA for visiting associateship. RB thanks Prof. Banibrata Mukhopadhyay, Department of Physics, IISC, Bangalore for regorous discussions regarding this problem earlier.


\newpage
\addcontentsline{toc}{part}{\bf  Bibliography}
\markright{ Bibliography}

\end{document}